\documentclass{vgtc}                          

\ifpdf
  \pdfoutput=1\relax                   
  \usepackage{graphicx}                
  \DeclareGraphicsExtensions{.pdf,.png,.jpg,.jpeg} 
\else
  \usepackage{graphicx}                
  \DeclareGraphicsExtensions{.eps}     
\fi%

\graphicspath{{figures/}{pictures/}{images/}{./}} 

\usepackage{microtype}                 
\PassOptionsToPackage{warn}{textcomp}  
\usepackage{textcomp}                  
\usepackage{mathptmx}                  
\usepackage{times}                     
\usepackage{cite}                      
\usepackage{tabu}                      
\usepackage{booktabs}                  
\usepackage{multirow}

\onlineid{0}

\vgtccategory{Research}

\vgtcinsertpkg

\title{Using Trajectory Compression Rate to Predict Changes in Cybersickness in Virtual Reality Games}

\author{Diego Monteiro\thanks{\textit{Corresponding author}; e-mail: Diego.VilelaMonteiro@bcu.ac.uk}\\ %
    \parbox{1.7in}{\scriptsize \centering Department of Computing \\ Xi’an Jiaotong-Liverpool University \\ DMT Lab \\ Birmingham City University}

\and Hai-Ning Liang\thanks{\textit{Corresponding author};  e-mail: HaiNing.Liang@xjtlu.edu.cn}, Xiaohang Tang\thanks{e-mail: Xiaohang.Tang19@student.xjtlu.edu.cn}\\ %
    \parbox{1.7in}{\scriptsize \centering Department of Computing \\ Xi’an Jiaotong-Liverpool University}

\and Pourang Irani\thanks{e-mail: Pourang.Irani@cs.umanitoba.ca}\\ %
    \parbox{1.7in}{\scriptsize \centering Department of Computer Science \\ University of Manitoba}}

\abstract{Identifying cybersickness in virtual reality (VR) applications such as games in a fast, precise, non-intrusive, and non-disruptive way remains challenging. Several factors can cause cybersickness, and their identification will help find its origins and prevent or minimize it. One such factor is virtual movement. Movement, whether physical or virtual, can be represented in different forms. One way to represent and store it is with a temporally annotated point sequence. Because a sequence is memory-consuming, it is often preferable to save it in a compressed form. Compression allows redundant data to be eliminated while still preserving changes in speed and direction. Since changes in direction and velocity in VR can be associated with cybersickness, changes in compression rate can likely indicate changes in cybersickness levels. In this research, we explore whether quantifying changes in virtual movement can be used to estimate variation in cybersickness levels of VR users. We investigate the correlation between changes in the compression rate of movement data in two VR games with changes in players' cybersickness levels captured during gameplay. Our results show (1) a clear correlation between changes in compression rate and cybersickness, and (2) that a machine learning approach can be used to identify these changes. Finally, results from a second experiment show that our approach is feasible for cybersickness inference in games and other VR applications that involve movement. %
} 

\CCScatlist{
  \CCScatTwelve{Human-centered computing}{Empirical studies in HCI}{}{};
  \CCScatTwelve{Human-centered computing}{Virtual reality}{}{}
  \CCScatTwelve{Human-centered computing}{HCI design and evaluation methods}{}{}
}

\begin{document}


\firstsection{Introduction}

\maketitle

Cybersickness affects many virtual reality (VR) users and remains a significant challenge \cite{ref1,ref2}. Although cybersickness has been studied widely \cite{ref1,ref2,ref3,ref4/ref12,ref5,ref6,ref7,ref8/ref17,ref9,ref10}, the underlying factors that can help identify and predict it are not yet fully understood \cite{ref11}. Identifying cybersickness is a challenging yet essential task if we want to minimize or eliminate it. As we identify and understand more of the factors associated with cybersickness \cite{ref4/ref12,ref13}, the easier it will be to create efficient, effective, and simple non-intrusive and non-disruptive ways to prevent or at least mitigate it \cite{ref4/ref12,ref5,ref14,ref15,ref16}.

One well-known trigger of cybersickness symptoms is the illusion of movement caused by visual stimuli without the accompanying vestibular ones (e.g., the feeling of motion when seeing an adjacent vehicle move while the observer is stationary) \cite{ref8/ref17}. Further, it has been shown that rotational movement, such as turning or looking sideways, is more sickening than translational movement (i.e., moving facing a constant direction) \cite{ref18,ref19,ref20,ref21}.

One way to determine cybersickness is to measure the type of movement users make, whether it has rotational changes or just unidirectional movement. To do this measurement, we need to represent movement computationally. One of the most common and intuitive ways to represent and measure movement data is through a sequential time series of positional points \cite{ref22}. One feature of such a representation is that it can be simplified and compressed \cite{ref23}. Data compression techniques work best on redundant (or repeated) data \cite{ref24,ref25,ref26}—that is, data that represent a path that is unidirectional than with many turns and rotation changes \cite{ref27}. It is possible to assume that movement patterns following a straighter path at a constant speed present greater redundancy and are more prone to compression. Accordingly, we can use the compression rate of movement data to detect whether users follow a straighter path or have many turns. That is, we can use compression rate as a measure of possible cybersickness experienced by VR users. 

Our literature review and understanding of compression rate led us to two hypotheses: \textit{(H1) There is a negative correlation between compression rate and cybersickness; and (H2) Changes in compression rate can be used to infer changes in cybersickness.} We ran an experiment where participants would move through two VR environments while having their discomfort and movement trajectories recorded periodically to evaluate these two hypotheses. After an in-depth analysis of the collected data, we were able to correlate changes in the compression of users' navigation patterns with changes in their cybersickness levels. We then applied our findings in a proof-of-concept experiment to predict the level of cybersickness in a navigation application. Our approach was able to predict with relatively little error the changes in cybersickness levels, both positively and negatively.

Our contributions are two-fold: (1) presenting a novel correlation between trajectory compression rate and cybersickness, and (2) demonstrating experimentally that a machine learning approach can be used to identify changes in cybersickness using this measurement.

This paper is organized as follows. We next present a literature review where we describe the theoretical background that led us to the two hypotheses mentioned above. We then present the environments and algorithms used to test and verify our two hypotheses, allowing us to create a functional prediction tool used in the second experiment. Finally, we discuss the results and their implications for future development.

\section{Theoretical background}
Researchers have been studying cybersickness for quite some time since even before the advent of current consumer head-mounted displays (HMD). However, its origins are still not entirely well understood \cite{ref1,ref28}. Studies point to a relationship with motion sickness. It has been empirically demonstrated that one factor associated with cybersickness symptoms is the \textit{illusion} of movement \cite{ref8/ref17}. However, it is difficult to pinpoint when these symptoms start because of the way cybersickness is measured. 

Another critical discussion is how to represent movement and identify changes in it, which can be used to predict cybersickness. Thus, this literature review is divided into two main themes: (1) measurements to identify cybersickness; and (2) computational representations of movement and its compression methods. They are then subdivided into four parts.

\subsection{Movement Type Affects Cybersickness}
Various behavioral and technological factors can influence the level of cybersickness experienced by people. These factors include having enough rest, practicing specific eating habits, and the screen's field of view and refreshing rate \cite{ref29}. Nevertheless, these factors can be challenging to observe and control from an application standpoint. Factors that can be registered within a VR application include self-motion components like velocity \cite{ref30/ref56} and rotational movement \cite{ref20}, which can be captured easily. 

Keshavarz et al. \cite{ref8/ref17,ref19} observed that self-motion alone is not enough to cause sickness but is a facilitator or enhancer because virtual velocity can increase cybersickness \cite{ref30/ref56}. Other studies found that the sickness level would increase just by adding a single type of rotation \cite{ref20}. Adding another type of rotation increased it even more \cite{ref19}. Thus, it is likely these kinds of movements (that is, changes in velocity and rotational navigation) performed by the user can be precursors of cybersickness. This observation could be one reason locomotion techniques that avoid or limit movement, such as Teleport, are found to mitigate cybersickness.

\subsection{Measuring Cybersickness}
The identification and measurement of cybersickness mostly rely on subjective metrics, such as questionnaires, like the Simulator Sickness Questionnaire (SSQ) \cite{ref31} and Virtual Reality Sickness Questionnaire (VRSQ) \cite{ref32} or marking the discomfort level in a scale midway through a game or after playing it \cite{ref33/ref58}. These approaches have clear disadvantages. Using questionnaires to measure cybersickness after completing the experiment loses the nuances and specificity needed to identify when and why participants are getting sick. Frequent incomplete information can potentially lead to erroneous conclusions and overgeneralizations. Besides, asking the participants to have a long pause or pauses mid-game and collect their sickness levels can be equally problematic outside of an experimental setting. It can potentially disrupt the immersive experience and be detrimental to the in-game flow experience and performance in games. In many fast-paced games, seconds between actions can be the difference between success and failure.

The alternatives to questionnaires or lengthy mid-game interruptions are physiological or psychophysiological metrics to detect cybersickness. These techniques usually rely on machine learning classification to analyze the data and have produced positive results \cite{ref34,ref35,ref36,ref37,ref38}. One notable result is that of Islam et al. \cite{ref34}, whose approach was able to predict cybersickness with 97\% accuracy. Other studies involve skin conductance that correlates to nausea measurements \cite{ref39} and eye movements that can explain over a third of the total variance in cybersickness \cite{ref40}. Further, these physiological techniques can be predictive, such as using data of eye movement \cite{ref40}, \cite{ref41}, heart rate \cite{ref42}, brainwaves \cite{ref43}, \cite{ref44}, and vestibular-evoked myogenic potentials \cite{ref45}. The approach presented in \cite{ref34} combined them and was able to achieve predictions with almost 90\% accuracy. Though all these techniques and approaches have potential, they present disadvantages. They often require expensive, cumbersome, and sometimes intrusive equipment, limiting the application potential of new techniques to mitigate cybersickness and making it difficult to assess sickness in commercial products outside a controlled research environment.

Some recent works have focused on information that can be acquired in-game rather than through expensive sensors \cite{ref46,ref47}. Porcino et al. \cite{ref46} use a series of in-game data (such as virtual movement) and data acquired from a demographics questionnaire to detect when participants start to have some form of discomfort. However, they relied on a long demographics questionnaire. Hell and Argyriou \cite{ref47} used a similar methodology to estimate the degree of nausea likely to be caused by a virtual roller coaster. However, their technique cannot work for dynamic, online detection.

In short, it would be of great advantage to be able to detect cybersickness fast and without resorting to expensive equipment or significantly disrupting immersion and flow during VR experiences with long pauses. As stated earlier, because movement in VR can be associated with sickness \cite{ref46}, it might be possible to use changes in movement during user-VR interaction as a simple and inexpensive marker to gauge alterations in discomfort and cybersickness. 

\subsection{Movement Representation}
Representing natural phenomena computationally is challenging \cite{ref48}. For every phenomenon, various ways can be used to capture and describe specific aspects or fit within the machine's limitations \cite{ref49,ref50}. One such phenomenon that can be represented in several ways is movement. Possible ways to represent it include vectors and flow networks \cite{ref51} or equations of motion \cite{ref52}. While these kinds of representations are helpful and can be memory efficient, they often require extensive pre-processing and are thus not adequate for dynamic applications \cite{ref23,ref52}.

Trajectories are one of the most appropriate ways of representing and storing movement \cite{ref53}. According to Renso et al. \cite{ref51}, trajectories are segments of movement. Although movement is inherently continuous, it cannot be stored as such in computers where the stored data is discrete. Alvares et al. \cite{ref22} have stated that data representing moving objects are ordinarily available as sample points in the form of (tid, x, y, z, t). In this representation, tid is an object identifier while x, y, z, and t are respectively a set of three-dimensional spatial coordinates and a time stamp. This kind of representation is commonly used in scientific and commercial applications \cite{ref27}. We chose to work with this discrete representation because it is the most common and more easily translatable to an active gaming environment or other similar VR applications. It can be used in real-time with little or no pre-processing because these stored trajectories can be simplified further by compression, reducing memory demand and processing power.

\subsection{Trajectory Compression} \label{session:2-4}
In 1973, Douglas-Peucker proposed one well-known compression technique, which has been studied and improved over the years \cite{ref54}. This technique calculates the distance between the current point and the path drawn by a straight line that is imagined between both the current point's neighbors. If the distance is below a threshold, the point is eliminated. It was designed for line simplification and, although it can be used to compress trajectories, it often affects the temporal aspects (such as stored object speed). 

To minimize the Douglas-Peucker algorithm's loss when dealing with spatiotemporal data, Meratnia and By \cite{ref54,ref55} have proposed a new technique that takes into consideration both speed and position when compressing trajectory data \cite{ref23}. Their technique works differently and, instead of considering if the middle point should be eliminated, this technique works with a pair of points at a time. Using the position and speed of the pair, they extrapolate the next point and, if the distance between the predicted and actual point is below a threshold, the following point becomes the last in the pair and the first remains the same; otherwise, a new pair is formed starting from the former last point. Because Meratnia and By's technique does not require knowledge of the whole setup, it can be used in dynamic applications \cite{ref54,ref55}. Although simple in conceptualization, compression approaches based on this technique are very effective and efficient, making them relevant to various real-life applications \cite{ref27}. 
 
\begin{figure}[tb]
 \centering 
 \includegraphics[width=\columnwidth]{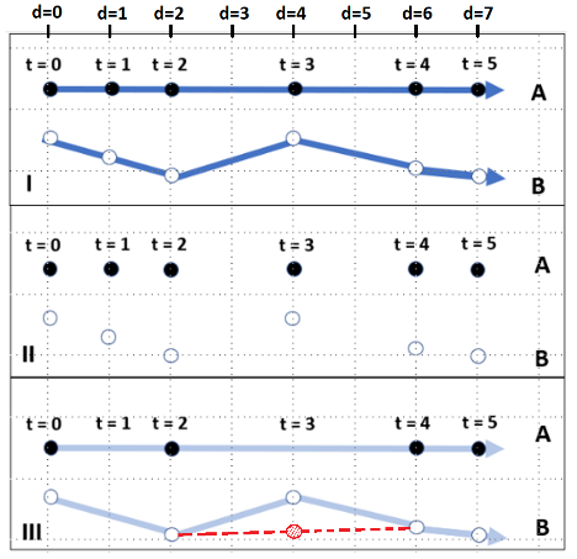}
 \caption{Example of trajectory sampling, storage, and compression. Time (t) and distance (d) are abstract values and constant in all figures. (I, Top) Trajectories A and B are sampled at regular intervals (t=1); both trajectories are faster between t=2 and t=4 than during the other intervals. (II, Middle) Later, they are stored as a sequence of points. (III, Bottom) Finally, the points that do not represent changes in direction or speed are removed. Trajectories A and B both lose a point at t=1, whose speeds and positions can be inferred from their respective t=0 and t=2. Only Trajectory A loses a point at t=3; its speed and position can be inferred from t=2 and t=4. In Trajectory B, the equivalent point represents a change in direction; if it was eliminated, an erroneous t=3 could be estimated (red striped path and circle).}
 \label{fig:figure1}
\end{figure}

\autoref{fig:figure1} shows two examples of trajectories that are represented and later compressed. Trajectories A and B are sampled at regular intervals (\autoref{fig:figure1}.I). Later, they are stored as a sequence of points (\autoref{fig:figure1}.II). Finally, the points that do not represent changes in direction or speed are removed. In this example, two points are removed from Trajectory A, whereas only one is removed from Trajectory B (\autoref{fig:figure1}.III).

Two observations are important to highlight. First, compression is the most effective when redundancy is present \cite{ref24,ref27}, for example, in a straight line or a stationary object. Second, turns and increases in speed have been demonstrated to increase sickness in virtual environments \cite{ref30/ref56}. These two observations led us to hypothesize that a trajectory with many redundant points, such as a straight line, will be less likely to cause sickness than a trajectory that cannot be compressed (or has a lower rate of compression) because of the presence of turns. More succinctly, we formed the two hypotheses presented in the introduction.

\section{Experiment A}
We conducted a two-part experiment to collect the necessary data for correlation analysis. The experiment aimed to investigate if the results would support our hypothesis of the correlation between cybersickness and trajectory compression and inform the design of the following experiment that would access its applicability in a gaming environment. 

This experiment was classified as low risk research and was approved by the University Ethics Committee at Xi’an Jiaotong-Liverpool University (\#21-01-09). 

\subsection{Participants}
We recruited a total of 10 participants (3 females) from a local university with an average age of 19.3 (s.d. = 0.46), ranging from 19 to 20. All volunteers declared to have normal or corrected-to-normal vision, and none of them declared any history of color blindness or health issues. Two had experience with VR systems. 

\subsection{Apparatus}
We used an Oculus Rift S with its accompanied controller as our HMD because it is one of the most common, popular ready-made VR devices. A desktop with 16GB RAM, an Intel Core i7-7700k CPU @ 4.20GHz, a GeForce GTX 1080Ti dedicated GPU, and a standard 21.5" 4K monitor were used to drive the HMD. 

\subsubsection{Game Environments}
The games in our experiment were developed in-house using Unity. We implemented them in-house to avoid or minimize confounding factors. We created the environment using the metric system and defined one Unity unit to be equal to one meter \cite{ref57}. To avoid confusion with real-world movement, we would refer this measurement unit as Unity Meter (Um for short). 

\begin{figure}[tb]
 \centering 
 \includegraphics[width=\columnwidth]{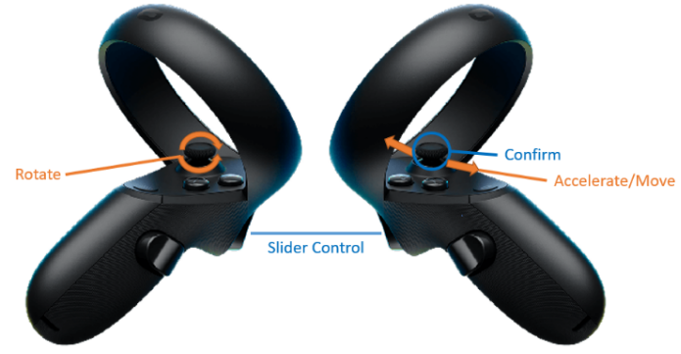}
 \caption{A diagram of the dual handheld controller and its functions. The orange markings show the controls associated with movement of the character. The left joystick is used to rotate the character’s viewpoint while the right joystick is used to move the character. The blue markings show the controls associated with entering the Discomfort Score (see also \autoref{fig:figure5} below). The two trigger buttons are used to move the slider left and right and pressing the right joystick is used make a confirmation. }
 \label{fig:figure2}
\end{figure}

The two games consisted of one maze escape (Maze) and the other was an obstacle race-like challenge (Race). They were representative of typical first-person tasks in which players were required to get to the end before running out of time. As such, all character movements (i.e., acceleration and rotation) were performed using the joysticks in the controller. They were designed to work with a standard Oculus controller (see \autoref{fig:figure2}). The movement of the character was controlled using the controller's joysticks and happened in a continuous fashion (standard Unity movement). In other words, pushing the joystick towards the desired direction would see all the frames relative to that movement.

\begin{figure}[tb]
 \centering 
 \includegraphics[width=\columnwidth]{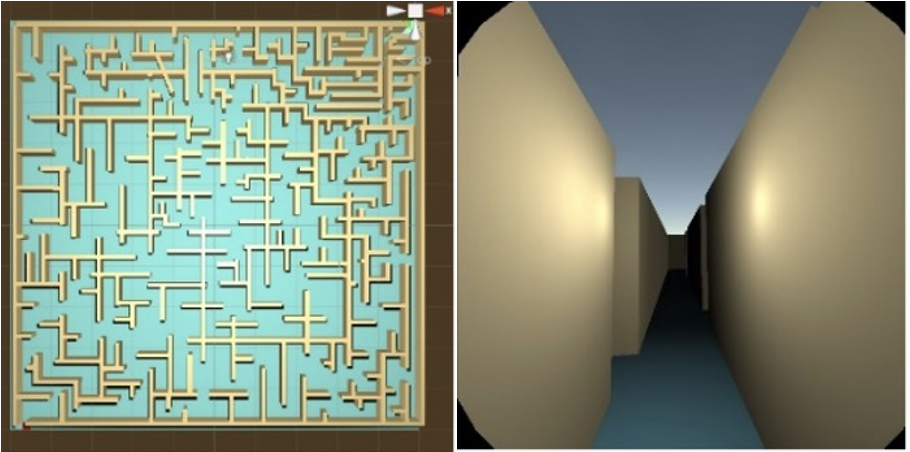}
 \caption{The maze environment used in our experiment. (Left) A bird's-eye view of the maze; it needs players to make at least 40 turns to find the exit. (Right) A view of the high wall in the maze from within.}
 \label{fig:figure3}
\end{figure}

The first game, the maze, was a typical escape game (task). Players were positioned at the beginning of the maze and instructed to find the exit within a specified time. The environment was open on the top and had tall beige walls towering beyond the player's field of view. The maze did not have any other visual cues that could lead to path memorization (see \autoref{fig:figure3}). It was designed to make players turn their characters (i.e., their camera) at least 40 times at 90º or more, a design that would increase rotation and thus the possibility of cybersickness \cite{ref18,ref19,ref20,ref21}. There were no other adversarial or stressing factors (e.g., no enemies, no path changes, no clock ticking noises), nor any other incentives to turn. During movement, their speed was 2 Unity Meters per second (Um/s). 

\begin{figure}[tb]
 \centering 
 \includegraphics[width=\columnwidth]{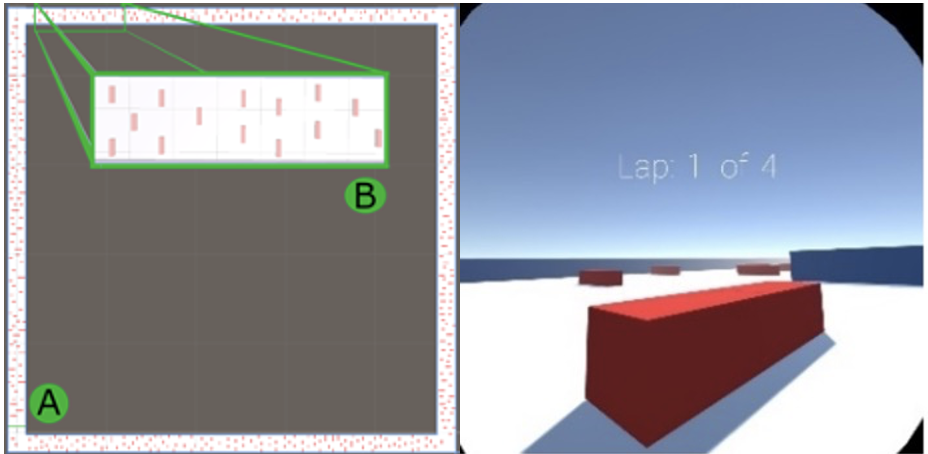}
 \caption{The race environment used in our experiment. (Left) A bird's-eye view of the track. (A) represents the start and finish points within the tracks; (B) shows a zoomed-in region of the track. (Right) A view of the Race game environment, where there is an obstacle that the user needs to avoid running into as seen from within the track.}
 \label{fig:figure4}
\end{figure}

The second game environment, the race, consisted of an obstacle course. Players were positioned at the starting point and informed they had to follow a roughly defined path (a lap) within the least possible time. In the path, there were barriers to be circumvented by the player. The obstacles were visibly colored red to avoid any confusion from the player (\autoref{fig:figure4}). Fog was used to hide the obstacles beyond five rows. In this task, the player could control the speed via the controller's joystick (with a maximum speed of 10 Um/s). 

\subsubsection{Evaluation Metrics}
\begin{figure}[tb]
 \centering 
 \includegraphics[width=\columnwidth]{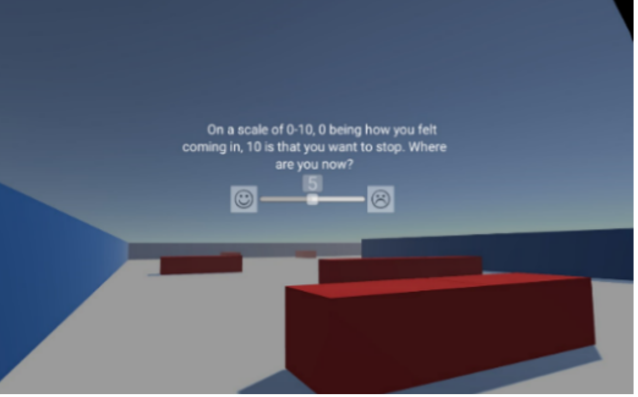}
 \caption{An example of a screenshot when the Discomfort score is being collected during the race game while it is still running in the background.}
 \label{fig:figure5}
\end{figure}

To collect the in-game Discomfort Score quickly and as non-interruptive as possible, we adopted the technique reported in \cite{ref33/ref58}. We presented on the screen a sliding Scale from 0 to 10 to collect the participants' current level of sickness dynamically while the game would still be running in the background (see \autoref{fig:figure5}). Zero meant no discomfort, and ten meant that the participant felt so ill that the game (and thus the experiment) had to be stopped. Participants used the back-triggers to move the slider to the left or right and pressed the right joystick to confirm the answer (see also \autoref{fig:figure2} above). Typically, it would take less than a second for participants to make a selection.

The participants' position was recorded using a logging function in Unity. In this function, the character's (camera's) global position in the world was recorded in the (tid, x, y, z, t) format. A sequence of these positions was then used in the dynamic compression algorithm. The algorithm used to compress the player's trajectory is based on the implementation of the spatiotemporal compression algorithm \cite{ref55} described in \cite{ref23} (see also \autoref{session:2-4}). It uses the speed and direction between two points to estimate where a third one would be. It would keep the points if either speed or direction changed. 

We used a threshold of 0.4 Unity Meters (Um) between real and expected positions to decide what points to eliminate. This threshold was selected through prior experimentation, where we found that such value represented a suitable balance and worked reliably in our context. The results from this experimentation pointed to a value between 0.25 Um and 0.5 Um. Smaller values would barely remove any points, while greater values would remove virtually all the points. In the end, we decided to use 0.4 Um for the current study. As our results would show, this worked well in our experiments.

We then evaluated the compression rate of the segment by dividing the eliminated points by the original number of points. We calculated the compression rate within 2-minute windows, which matched the Discomfort Score measurements. The changes in compression rate were calculated based on the difference between the current $x$-minute window compression rate and the previous $x$-minute window compression rate, where $x$ is a time pre-determined and can be adjusted when needed. In other words, let $Rw$ be the number of points removed from the trajectory by the compression algorithm in the current $x$-minute window $w$, and $Tw$ be the total number of points in the same window $w$. We then have the compression rate $Cw$ as Equation 1:

\begin{equation}
C_w = R_w / T_w
\end{equation}

Then, we can define the compression rate in the former $x$-minute window $w-1$ as Equation 2:

\begin{equation}
C_{w-1} = R_{w-1} / T_{w-1}
\end{equation}

Finally, we have the difference in compression rate between $x$-minute windows as Equation 3:

\begin{equation}
\Delta C_w = C_w / C_{w-1}
\end{equation}

\subsection{Procedure}
Experiment A was divided into two tasks, one for each game, to collect data from participants during the two different trajectory scenarios.

Upon arrival and after consenting to participate in the experiment, all participants completed a demographics questionnaire and were divided into two groups, A and B. They were then instructed to play the game according to their group. Group A first played the maze, which was slower but with more turns, followed by at least a 15-minute rest, and then they would play the race. Group B was presented with the games in reverse order. Participants could rest longer than 15 minutes if they wanted to. In all conditions, the participants were sitting upright. The chair did not rotate or rock.
In both games, participants were informed that, although the challenge was to finish as fast as possible, they could stop moving in the virtual world if they felt any discomfort. They were allowed to pause or stop the game and ask questions about the experiment at any time if they so desired. 
Participants could quit the experiment if they so desired. Because of the nature of the experiment, data from participants who left the experiment earlier were still useful. Moreover, pauses were also logged because they affected the compression rate.

Due to Covid-19, and although the situation was considered quite safe to run experiments within campus, participants and researchers were required to wear a mask and follow other additional safety protocols to reduce any possible spread of the virus further.

\subsection{Results}
We used both statistical inference methods and visualizations to analyze the data. A Shapiro-Wilk test was conducted first to check for normality of data. We used parametric tests for normally distributed data; otherwise, we used non-parametric tests.

We conducted Mauchly's Test of Sphericity for normally distributed data and employed Repeated Measures ANOVA (RM-ANOVA) with Bonferroni correction to detect overall significant differences. We applied the Greenhouse-Geisser correction when faced with a violation of the assumption of sphericity. We used Pearson's correlation for normally distributed data; otherwise, Spearman's correlation was used instead.

\subsubsection{Discomfort Score}
\begin{figure}[tb]
 \centering 
 \includegraphics[width=\columnwidth]{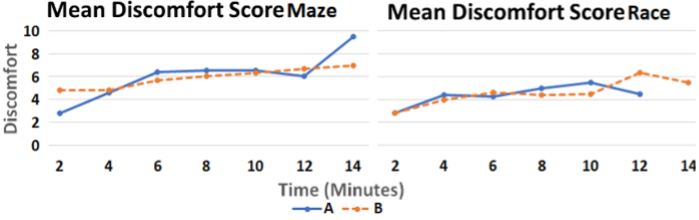}
 \caption{Mean Discomfort Scores according to the two groups for the Maze and the Race games.}
 \label{fig:figure6}
\end{figure}

On average, the mean Discomfort Score increased 1.5 points every 2 minutes in the Maze game for Group A, and 0.4 points every 2 minutes for Group B (see \autoref{fig:figure6}). The mean Discomfort Score change in the Race condition was 0.3 for Group A and 0.45 for Group B.

Group A had one participant who reached a Discomfort Score of 10 and quit the Race game before 6 minutes. All other participants were able to complete the Race game (8 before the 14-minute mark). The Maze game had a greater early termination rate due to some participants reaching an elevated level of discomfort; three participants from Group A (minute 10) and two from Group B (minute 6).

The final Discomfort Score between participants who played the Maze game first and the those who played the Race game first was significantly different after the first game ($F(1,8) = 29.333$, $p < .001$). However, the final Discomfort Scores after having played both were not statistically different between groups ($F(1,8) =.027$, $p = .874$).

\subsubsection{Compression Rate}
The compression rate did not vary between the two groups, but it did vary between game conditions—this is a good indication that compression rate can help predict changes in navigation patterns. The compression rate for the Maze conditions was higher than the one for the Race conditions. On average, the algorithm eliminated 35.5\% of trajectory points in the Race game and 77.6\% of the trajectory points in the Maze game (that is, a difference of 42\%). 

Although the compression rate was quite different in both conditions, the mean variation between subsequent compression rate was similar (-1.2 in the Race condition and -1.1 in the Maze condition) and was further validated by an RM-ANOVA ($p = .844$). The absolute compression rate is weakly correlated to the absolute Discomfort Score ($r_s(103) = .379$, $p < .001$). However, the variation in the compression rate and the variation in the Discomfort Score were moderately correlated ($r_s(81) = .492$, $p < .001$). 

\subsubsection{Predicting the Discomfort Score}
Using the standard Multi-Layer Perceptron (MLP) function from SPSS 24\footnote{https://www.ibm.com/analytics/spss-statistics-software}, the compression rate, and changes in compression as input parameters, it is possible to determine if the Discomfort Score would rise in over 90\% of cases. However, the MLP struggled to predict if the discomfort would be constant (see \autoref{tab:table1}). 

\begin{table}[tb]
  \caption{Prediction against Classification. Lower means that the Discomfort Score would decrease, and higher means that it would increase. The empty row in between means no change would happen.}
  \label{tab:table1}
  \scriptsize%
	\centering%
  \begin{tabu}{cccccc}
  \toprule
  \multicolumn{6}{c}{Classification}\\
  \midrule
  \multicolumn{2}{c}{\multirow{2}{*}{Sample}}&\multicolumn{4}{c}{Predicted}\\
  \cmidrule{3-6}
  &&Lower&&Higher&Correct\\
  \midrule
  \multirow{4}{*}{Training}&Lower&7&1&6&50.0\%\\
  \cmidrule{2-6}
  &&0&4&6&40.0\%\\
  \cmidrule{2-6}
  &Higher&1&1&33&94.3\%\\
  \cmidrule{2-6}
  &Global&&&&\textbf{74.6\%}\\
  \midrule
  \multirow{4}{*}{Test}&Lower&5&0&1&83.3\%\\
  \cmidrule{2-6}
  &&0&1&1&50.0\%\\
  \cmidrule{2-6}
  &Higher&0&1&13&92.9\%\\
  \cmidrule{2-6}
  &Global&&&&\textbf{86.4\%}\\
  \bottomrule
  \end{tabu}%
\end{table}

For this model, we used a 70/30 division between training and testing with normalized data. The MLP had one hidden layer with three neurons and the SoftMax activation function (the standard in SPSS 24). Both compression rate and changes in compression rate were equally important for the predictions. 

\subsection{Discussion}
Our primary research question has been answered negatively. Our results do not suggest a negative correlation between compression and Discomfort Score (\textit{H1}). However, our analysis indicates that it is indeed possible to use changes in the trajectory's compression rate to identify changes in the level of cybersickness (\textit{H2}). Our results show that because of the way that compression changed, it is quite possible to use the variations in compression rate as one of the reference factors for identifying cybersickness dynamically during gameplay. 

Moreover, our results show that this technique can be applied to other environments that use the same kind of controller and movement because it is not based on a single set of parameters but on the user's current changes in behavior pertinent to that environment. It worked quite well regardless of the order of the two games played by the participants. As such, this could be an inexpensive technique to detect cybersickness levels periodically without human intervention or expensive sensing devices. Also, it could help identify and provide visual or haptic mitigation techniques to be included in the VR application or just to suggest stopping times for players.

Although the compression technique had difficulties predicting when the discomfort remained stable, this could be expected from our data. The majority of our data represented either increases or decreases in the Discomfort Score, making their detection more precise. Moreover, a participant might interpret a slight increase in discomfort as an unchanged discomfort from the previous measurement. In contrast, the same participant might consider an unchanged discomfort as a slight increase or decrease after a while.

Finally, regarding the higher compression rate in the Maze compared to the Race, we hypothesize that this is because in the race game, players did smaller turns more often. As shown in \autoref{fig:figure1}, turns do not have to be complete to be a change in trajectory. Moreover, because participants had more flexibility to control the speed with the joystick, there was the need for more points to be kept by the algorithm in order to preserve the trajectory. On the other hand, in the maze game, the turns and speed were more controlled, creating straighter paths that could be more easily compressed.

\section{Experiment B}
\begin{figure}[tb]
 \centering 
 \includegraphics[width=\columnwidth]{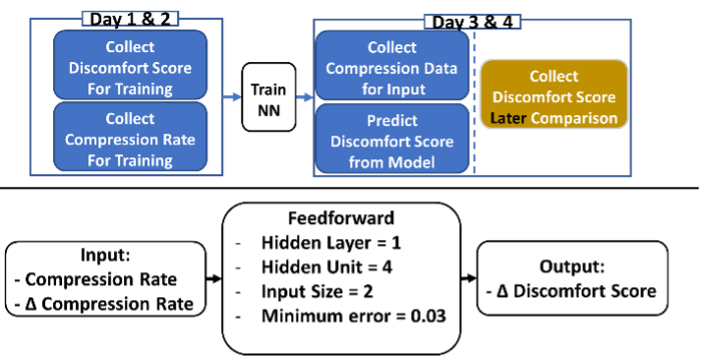}
 \caption{A high-level architecture diagram of the experiment (upper part) and feedforward NN (lower part). In the first two days, data is collected and used to train the NN. During days 3 and 4, the NN was tested, and the Discomfort Scores were collected only for our reference. The NN is presented in black and white with its data and parameters.}
 \label{fig:figure7}
\end{figure}

After showing high accuracy in detecting increases and decreases in the Discomfort Score from the MLP in Experiment A, we decided to implement a proof-of-concept neural network (NN) to predict changes in the Discomfort Score of individual users in an application periodically. We conducted a four-day experiment to collect the NN training data (for two days) and test it (in the last two days), as presented in the upper part of \autoref{fig:figure7}. 

\subsection{Participants}
We recruited a total of 15 participants (2 females) from the same local university. They had an average age of 22.47 (s.d. = 6.95), ranging from 20 to 47. All declared that they had normal or normal-to-corrected vision, and none declared any history of color blindness or health issues. Seven participants had experience with VR systems. 

On the last day, one participant (female, 47) quit the study for personal reasons, resulting in participants being divided equally among the two new navigation environments. The new average age was then 20.79 (s.d. = 1.42), now ranging from 20 to 25.

\subsection{Apparatus}
We used the same equipment as in Experiment A: An Oculus Rift S as our HMD and a desktop with 16GB RAM, an Intel Core, and a dedicated GPU.

\subsubsection{Neural Network}
The NN was a simple feedforward NN developed by a third party for Unity and available on GitHub\footnote{github.com/Blueteak/Unity-Neural-Network}. The model took as input both the current compression rate and its difference from the previous compression rate. Both were scaled to fit between 0 and 1. It had one hidden layer with four neurons, and the output was the variation in Discomfort Score. It was also scaled to fit between 0 and 1 (see \autoref{fig:figure7}, lower part). We used a 70/30 division of data between training and testing for each participant. That is, the NN was trained with each participant's data separately, and each individual result was used to predict the Discomfort Score of its respective participant.

\subsubsection{Navigation Environments}
\begin{figure}[tb]
 \centering 
 \includegraphics[width=\columnwidth]{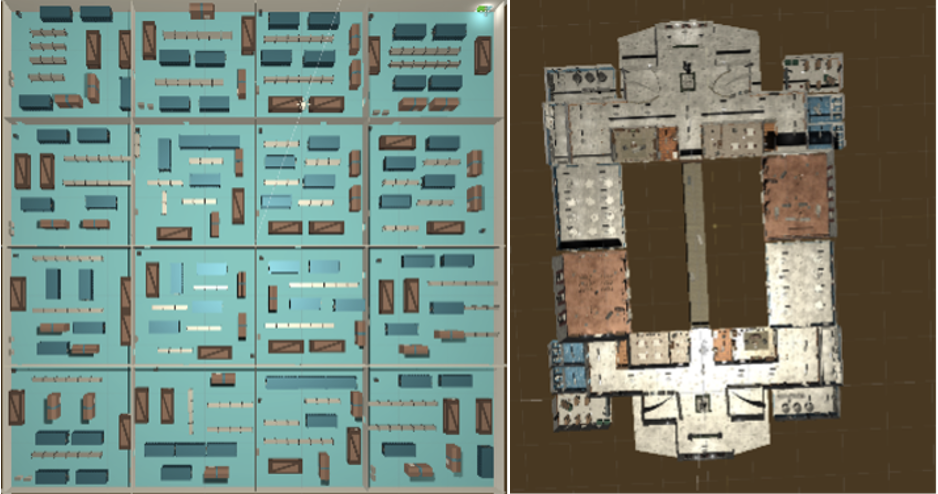}
 \caption{An aerial view of the environments used on the fourth day. (Left) Storage unit with shelves and boxes. (Right) Mansion with different rooms. Participants were instructed to walk inside them to look for an exit.}
 \label{fig:figure8}
\end{figure}

On the first three days of the experiment, all participants were asked to play the maze game (the same as in Experiment A). On the fourth (last) day, participants were randomly assigned to two different navigation environments and asked to navigate freely in them (see \autoref{fig:figure8}). These different environments had been developed in Unity and used the same type of control (continuous motion using only the joysticks on the controller). However, the environments presented different paths, objects, and lighting compared to the first experiment.

\subsection{Procedure}
As in the first experiment, participants answered a demographics questionnaire and received the general instructions. They then played the environments for 15 minutes each day. As they played the game, their positions were logged twice per second in the form (x, y, z, t) in the same way as in the previous experiment. Furthermore, their Discomfort Score and respective compression rate were registered every minute (that is, twice as often compared to the first experiment). We collected the data more often to have more data to train the model. All other procedures were the same as in the first experiment. The data on the first two days were used for training the NN.

On the third and fourth (i.e., the last two) days the data collected was used only to estimate the Discomfort Score. That is, it was not used in any way to train the NN. Each trained NN was participant-specific.

The prediction was calculated in the same periods that the participant would input their Discomfort Score. Note that the participants' discomfort was not used in the calculation, nor were the participants aware that a prediction existed. 

\subsection{Results}
We used both statistical inference methods and visualizations to analyze the data. A Shapiro-Wilk test was conducted first to check for normality of the data. We used parametric tests for normally distributed data; otherwise, non-parametric tests were used instead. Moreover, to calculate the error between the two graphs, we computed the difference between curves, an integrative method that accounts for mismatched peaks and valleys. For this calculation, the area between the reported and predicted scores is defined and divided by the reported score area.

\begin{figure}[tb]
 \centering 
 \includegraphics[width=\columnwidth]{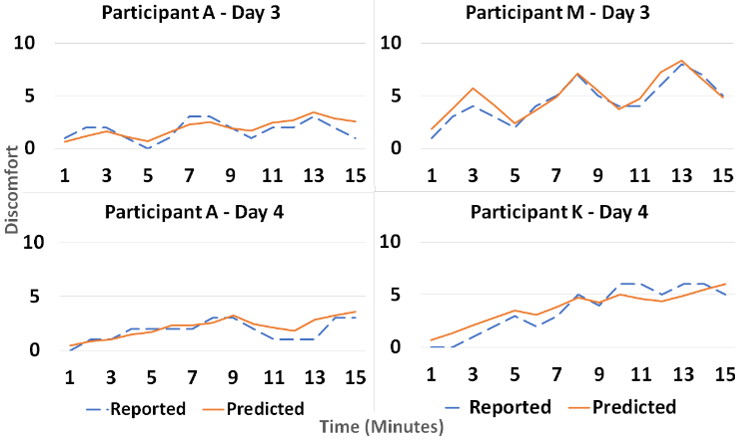}
 \caption{Discomfort Score over time of three participants during the 15-minute span of the experiment in Day 3 (upper part) and Day 4 (lower part). The dashed blue lines represent the Discomfort Score reported by each participant, while the continuous orange lines are the predictions. The X-axis represents time in minutes, and the Y-axis the Discomfort Score. The Discomfort Score was predicted well for both days for the same participants (i.e., Participant A) and different ones (i.e., Participants M and K).}
 \label{fig:figure9}
\end{figure}

\begin{figure}[tb]
 \centering 
 \includegraphics[width=\columnwidth]{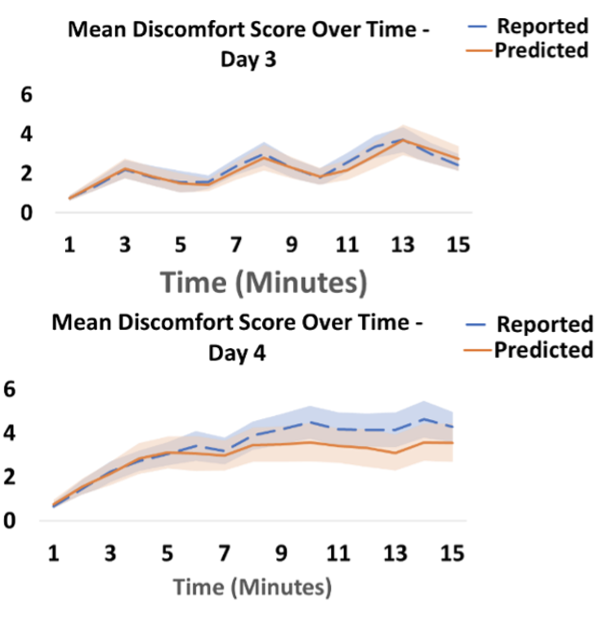}
 \caption{Mean Discomfort Score over time of all the participants over the 15-minute span of the experiment in Day 3 (upper part) and Day 4 (lower part). The dashed blue line represents the mean of the Discomfort Score reported by all the participants, while the continuous orange line is the mean of the predictions. The X-axis represents time in minutes, the Y-axis the Discomfort Score, and the shaded background the 95\% confidence interval.}
 \label{fig:figure10}
\end{figure}

Because the data was not normally distributed, we used the Wilcoxon signed-rank test to compare the predicted and reported Discomfort Score. This test did not find statistically significant differences between the reported and predicted scores in neither the third nor fourth days ($Z = -.739$, $p = .460$ and $Z = -.925$, $p = .355$, respectively). We then calculated the correlations between the reported and predicted scores using Spearman's $\rho$ correlation, which were identified as strong for both the third and fourth days ($r = .753$, $p < .001$ and $r = .608$, $p < .001$, respectively). Finally, the error (the difference between curves) concerning the mean reported values and the mean predicted values calculated by their areas was 2.9\%. The mean difference between points was 0.04 Discomfort Units on the third day (see \autoref{fig:figure9} and \autoref{fig:figure10}-upper part). Finally, the error was 6.1\% on day four, with a mean difference between points of 0.5 Discomfort Units (see \autoref{fig:figure9} and \autoref{fig:figure10}-lower part).

\subsection{Discussion}
Our results indicate that our technique is an efficient, low-cost, and practical way to detect cybersickness for an individual user in situations that require repeated exposure to the same VR environment or similar tasks. The mean difference between predicted and reported cybersickness levels is remarkably positive. It is below what can be input into the traditional Discomfort Score (slider with integers). Our predicted Discomfort Score is continuous, which could be used for further refined predictions in the long run. For example, it could be used for more frequent predictions and produce intermediary values based on each user's needs.

Given that our technique does not significantly interrupt the navigation activity in the VR environment, it can be used for games or similar applications with minimal effect on users' experience compared to other techniques. Moreover, as it is a software-based technique without additional hardware, it is scalable, low-cost, and can be implemented in different VR HMDs. For long-term exposure, recalibration might be necessary as some people become immune to cybersickness. Nevertheless, we do not see it as a problem since the calibration data can happen dynamically during interaction or gameplay. The Discomfort Score can be based on a starting and ending score or during other natural pauses (for example, during respawn or toilet breaks) rather than a repeated interruption of the player activity. The development of such recalibration is beyond the scope of this investigation and could represent an avenue for further research.

Due to the comparative nature of our work, we had to collect players' Discomfort Score data during gameplay. We kept this as non-disruptive as possible and allowed the game to continue playing in the background. Users could effectively complete it within less than a second each time. In addition, collecting this data is not a concern in a real-life application since the predictions are independent of the reported Discomfort Score and can be performed based on the movement data alone. The traditional Discomfort Score measurement is not a recommended long-term measure for cybersickness in commercial gaming applications if enjoyment, immersion, or the outcome of the games can be affected. In this research, the data was collected because we needed a reference point. Furthermore, given that we were not evaluating immersion or enjoyment in our experiment, using it did not affect our collected data—participants in our experiments did not express any issues with the data collection procedure.

\subsection{Limitations and Future Work}
This study evaluated our results using the threshold of 0.4 Um, which we found empirically to be an appropriate value before our study. We measured the Discomfort Score every two minutes. This period was not entirely arbitrary, and a value had to be defined for training this out-of-the-box NN, in which the input and output were fixed-size vectors. Further studies can be done to identify the optimal configurations of collection time, other input parameters for the NN, other machine learning models, and any relation to other questionnaires (e.g., SSQ and VRSQ). Such in-depth, extended evaluations are interesting but go beyond the scope of this paper, whose goal was to explore if the compression rate could feasibly be used to detect changes in discomfort—we have achieved this in our experiments.

The pandemic affected somewhat the recruitment of participants and limited the sample size of the experiments. Because of the size, we follow the recommendation to control for both Type I and II errors; thus, we presented the p values more consistently \cite{ref59,ref60}. Nevertheless, our experiments' size falls within the normal range for HCI empirical studies like ours that are published in rigorously peer-reviewed venues \cite{ref61}. As shown earlier, the results we have gathered are promising and reliable.

Finally, though the compression code had been extensively tested in other applications, in this study, we did not evaluate the accuracy of the compression techniques since this was not the paper's goal and could be another possible line of future work.

\section{Conclusion}
In this work, we have explored if movement trajectory compression rate could be used as one marker to help identify cybersickness in virtual reality (VR) applications, especially games. Our results show that it is indeed possible to use the compression rate as a potential marker or indicator for cybersickness. Our findings also show that it alone does not account for all the discomfort variations. Nevertheless, using compression rate as input for a simple neural network makes it possible to obtain accurate predictions for when increases or decreases in discomfort levels would occur during gameplay. In this paper, we presented results from one experiment to show that compression rate is a strong determinant of possible levels of cybersickness. Overall, this approach is novel, simple, and does not require special hardware. Compared to other techniques, it does not disrupt the player's immersive and flow experience during gameplay because data can be collected dynamically while the game is still running. 

\acknowledgments{
The authors thank the participants for their time and the reviewers for their insightful comments that have helped improve the paper. The work was supported in part by Xi’an Jiaotong-Liverpool University (XJTLU) Key Special Fund (\#KSF-A-03).}

\bibliographystyle{abbrv-doi}

\bibliography{template}
\end{document}